\documentclass[twocolumn,showpacs,preprintnumbers,superscriptaddress,prb,aps,psfig]{revtex4}

\usepackage{graphicx}
\usepackage{epstopdf}
\usepackage{float}
\usepackage{color}
\usepackage{amsmath}

\begin{document}

\title{Coexistence of antiferromagnetic and ferromagnetic spin correlations in Ca(Fe$_{1-x}$Co$_x$)$_2$As$_2$ revealed by $^{75}$As nuclear magnetic resonance}

\author{J.~Cui}
\affiliation{Ames Laboratory, U.S. DOE, Iowa State University, Ames, IA 50011, USA}
\affiliation{Department of Chemistry, Iowa State University, Ames, Iowa 50011, USA}
\author{P.~Wiecki}
\affiliation{Ames Laboratory, U.S. DOE, Iowa State University, Ames, IA 50011, USA}
\affiliation{Department of Physics and Astronomy, Iowa State University, Ames, Iowa 50011, USA}
\author{S.~Ran*}
\affiliation{Ames Laboratory, U.S. DOE, Iowa State University, Ames, IA 50011, USA}
\affiliation{Department of Physics and Astronomy, Iowa State University, Ames, Iowa 50011, USA}
\author{S.~L.~Bud'ko}
\affiliation{Ames Laboratory, U.S. DOE, Iowa State University, Ames, IA 50011, USA}
\affiliation{Department of Physics and Astronomy, Iowa State University, Ames, Iowa 50011, USA}
\author{P.~C.~Canfield}
\affiliation{Ames Laboratory, U.S. DOE, Iowa State University, Ames, IA 50011, USA}
\affiliation{Department of Physics and Astronomy, Iowa State University, Ames, Iowa 50011, USA}
\author{Y.~Furukawa}
\affiliation{Ames Laboratory, U.S. DOE, Iowa State University, Ames, IA 50011, USA}
\affiliation{Department of Physics and Astronomy, Iowa State University, Ames, Iowa 50011, USA}

\date{\today}

\begin{abstract} 
    Recent nuclear magnetic resonance (NMR) measurements revealed the coexistence of stripe-type antiferromagnetic (AFM) and ferromagnetic (FM) spin correlations in both the hole- and electron-doped BaFe$_2$As$_2$ families of iron-pnictide superconductors by a Korringa ratio analysis.
    Motivated by the NMR work, we investigate the possible existence of FM fluctuations in another iron pnictide superconducting family, Ca(Fe$_{1-x}$Co$_x$)$_2$As$_2$. 
   We re-analyzed our previously reported data in terms of the Korringa ratio and 
found clear evidence for the coexistence of stripe-type AFM  and FM spin correlations in the electron-doped CaFe$_2$As$_2$ system. 
    These NMR data indicate that FM fluctuations exist in general in iron-pnictide superconducting families and thus must be included  to capture the phenomenology of the iron pnictides. 
 
\end{abstract}

\maketitle

   \section{Introduction} 
   Since the discovery of high $T_{\rm c}$ superconductivity in iron pnictides,\cite{Kamihara2008}  the interplay between spin fluctuations and the unconventional nature of superconductivity (SC) has been attracting much interest. 
    In most of the Fe pnictide superconductors, the ``parent" materials exhibit antiferromagnetic ordering below the N\'eel temperature.\cite{Canfield2010,Johnston2010,Stewart2011}
    SC in these compounds emerges upon suppression of the stripe-type antiferromagnetic (AFM) phase by application of pressure and/or chemical substitution, where the AFM spin fluctuations are still strong. 
    Therefore, it is believed that stripe-type AFM spin fluctuations play an important role in driving the SC in the iron-based superconductors, although orbital fluctuations are also pointed out  to be important. \cite{Kim2013}

   Recently nuclear magnetic resonance (NMR) measurements revealed that ferromagnetic (FM) correlations also play an important role in both the hole- and electron-doped BaFe$_2$As$_2$ families of iron-pnictide superconductors. \cite{Johnston2010,Wiecki2015, Wiecki2015prl}
            The FM fluctuations are found to be strongest in the maximally-doped BaCo$_2$As$_2$ and KFe$_2$As$_2$, but are still present in the BaFe$_2$As$_2$ parent compound, consistent with its enhanced magneric susceptibility $\chi$. \cite{Johnston2010}
     These FM fluctuations are suggested to compete with superconductivity and are a crucial ingredient to understand the variation of  $T_{\rm c}$ and the shape of the SC dome. \cite{Wiecki2015prl}
    It is interesting and important to explore whether or not similar FM correlations exist in other iron pnictide systems.

   The CaFe$_2$As$_2$ family has a phase diagram distinct from that for the BaFe$_2$As$_2$ family.
  Whereas for the BaFe$_2$As$_2$ materials the AFM and orthorhombic phase transitions become second order with Co substitution, the CaFe$_2$As$_2$ family continues to manifest a strongly first order, coupled, structural-magnetic phase transition even as Co substitution suppresses the transition temperature to zero.
   Another significant difference in the phase diagrams of the CaFe$_2$As$_2$ and BaFe$_2$As$_2$ systems is also found in superconducting phase.  
     Although SC appears when the stripe-type AFM phase is suppressed by Co substitution for Fe in both cases, no coexistence of SC and AFM has been observed in Ca(Fe$_{1-x}$Co$_x$)$_2$As$_2$, whereas the coexistence has been reported in Ba(Fe$_{1-x}$Co$_x$)$_2$As$_2$. 
    These results are consistent with the difference between a strongly first order versus second order phase transition.
    Recent NMR measurements revealed that the stripe-type AFM fluctuations are strongly suppressed in the AFM state in the Co-doped CaFe$_2$As$_2$ system, whereas sizable stripe-type AFM spin fluctuations still remain in the AFM state in the Co-doped BaFe$_2$As$_2$ system.\cite{Cui2015}
    These results indicate that the residual AFM spin fluctuations play an important role for the coexistence of AFM and SC in Ba(Fe$_{1-x}$Co$_x$)$_2$As$_2$.
    Furthermore, in the case of Ca(Fe$_{1-x}$Co$_x$)$_2$As$_2$, pseudogap-like behavior\cite{Cui2015} has been observed in the temperature dependence of 1/$T_1T$ and in-plane resistivity.  
    The characteristic temperature of the pseudogap was reported to be nearly independent of Co substitution. 

   In this paper, we investigated the possible existence of FM fluctuations in  Ca(Fe$_{1-x}$Co$_x$)$_2$As$_2$ and found the clear  evidence of coexistence of stripe-type AFM and FM correlations based on $^{75}$As NMR data analysis.
    In contrast to the case of Ba(Fe$_{1-x}$Co$_x$)$_2$As$_2$ where the relative strength of FM correlations increases with Co substitution, that of the FM correlations are almost independent of the Co content in  Ca(Fe$_{1-x}$Co$_x$)$_2$As$_2$ from $x$ = 0 to 0.059.  
   Although we have investigated a relatively small Co substitution region,  the existence of the FM spin correlations would be consistent with the fact that CaCo$_2$As$_2$, the end member of the electron doped Ca(Fe$_{1-x}$Co$_x$)$_2$As$_2$ family of compounds, has an A-type antiferromagnetic ordered state below $T_{\rm N}$  = 52--76 K\cite{Quirinale2013, Cheng2012} where the Co moments within the CoAs layer are ferromagnetically aligned along the $c$ axis and the moments in adjacent layers are aligned antiferromagnetically.
   Since the coexistence of FM and AFM spin correlations are observed in both the hole- and electron-doped BaFe$_2$As$_2$ systems,\cite{Wiecki2015prl} our results suggest that the FM fluctuations exist  in general in iron pnictide superconductors, indicating that theoretical microscopic models should include FM correlations to reveal the feature of the iron pnictides.

 \section{Experimental}
      The single crystals of Ca(Fe$_{1-x}$Co$_x$)$_2$As$_2$  ($x$ = 0, 0.023, 0.028, 0.033 and 0.059) used in the present study are from the same batches as reported in Ref. \onlinecite{Cui2015}. 
      These single crystals were grown out of a FeAs/CoAs flux,\cite{Ran2011,Ran2012} using conventional high temperature growth techniques.\cite{Canfield_book, Canfield_1992}
   Subsequent to growth, the single crystals were annealed at $T_{\rm a}$ =  350~$^{\circ}$C for 7 days and then quenched.  
   For $x$ = 0, the single crystal was annealed at $T_{\rm a}$ = 400~$^{\circ}$C for 24 hours. 
     Details of the growth, annealing and quenching procedures have been reported in Refs.~\onlinecite{Ran2011} and \onlinecite{Ran2012}.  
    The stripe-type AFM states have been reported below the N\'eel temperatures $T_{\rm N}$ = 170, 106, and 53 K for $x$ = 0,  0.023, and 0.028, respectively.\cite{Goldman2008}
    The superconducting states are observed below the transition temperature of $T_{\rm c}$ = 15 and 10 K for $x$ = 0.033 and 0.059, respectively.\cite{Ran2012}

    NMR measurements were carried out on $^{75}$As  (\textit{I} = 3/2, $\gamma/2\pi$ = 7.2919 MHz/T, $Q$ =  0.29 Barns)  by using a lab-built, phase-coherent, spin-echo pulse spectrometer.  
   The $^{75}$As-NMR spectra were obtained at a fixed frequency $f$ = 53 MHz by sweeping the magnetic field.
  The magnetic field was applied parallel to either the crystal $c$ axis or the $ab$ plane where the direction of the magnetic field within the $ab$ plane was not controlled. 
    The $^{75}$As 1/$T_{\rm 1}$ was measured with a recovery method using a single $\pi$/2 saturation $rf$ pulse. 
     Most of NMR experimental results were published elsewhere.\cite{Furukawa2014,Cui2015} 

\begin{figure}[tb]
\includegraphics[width=8.0 cm]{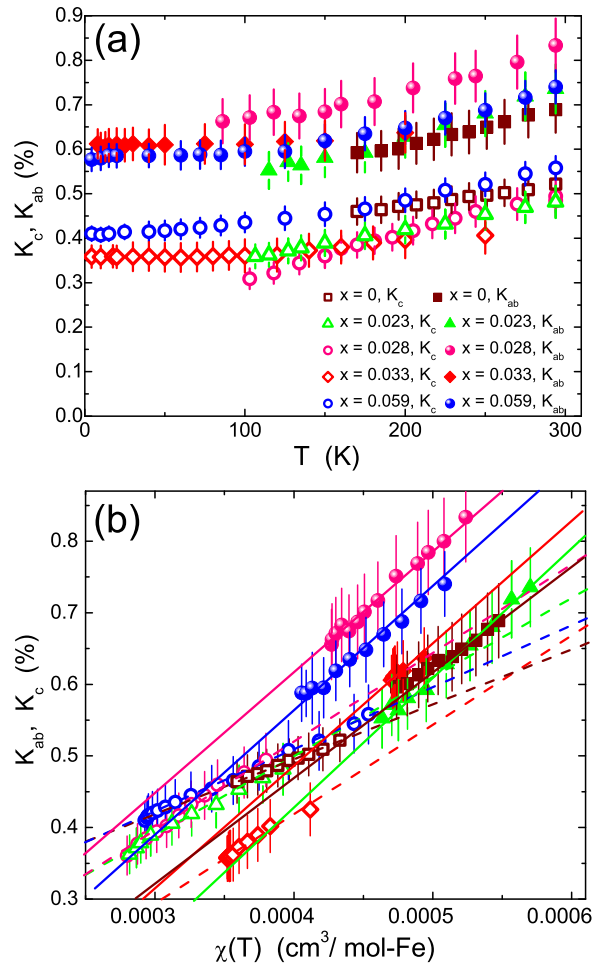} 
\caption{(Color online) (a) Temperature 
dependence of $^{75}$As NMR shifts $K_{ab}$ and $K_{c}$ for Ca(Fe$_{1-x}$Co$_x$)$_2$As$_2$. 
 (b) $K(T)$ versus magnetic susceptibility $\chi(T)$ plots for the corresponding $ab$ and $c$ components of $K$ in Ca(Fe$_{1-x}$Co$_x$)$_2$As$_2$ with $T$ as an implicit parameter. 
 The solid and broken lines are linear fits.
}
\label{fig:T-K}
\end{figure}

 \section{Results and discussion}

  In this paper we discuss magnetic correlations in Ca(Fe$_{1-x}$Co$_x$)$_2$As$_2$ based on a Korringa ratio analysis of the NMR results.  
   Figure \ \ref{fig:T-K}(a)  shows the $x$ and $T$ dependence of the Knight shifts, $K_{\rm ab}$ for $H$  parallel to the $ab$ plane and $K_{\rm c}$ for $H$  parallel to the $c$ axis, where new Knight shift data for $x$ = 0.033 and 0.059 are plotted in addition to the data ($x$ =0, 0.023 and 0.028) reported previously.\cite{Furukawa2014,Cui2015} 
     The NMR shift consists of a $T$-independent orbital shift $K_0$ and a $T$-dependent spin shift $K_{\text{spin}}(T)$ due to the uniform magnetic spin susceptibility $\chi(\mathbf{q}=0)$ of the electron system. 
     The NMR shift can therefore be expressed as $K(T)=K_0+K_{\text{spin}}(T)=K_0+A_{\text{hf}}\chi_{\text{spin}}/N$, where $N$ is Avogadro's number, and 
$A_{\text{hf}}$ is the hyperfine coupling constant, usually expressed in units of T$/\mu_{\rm B}$.
     Since detailed analysis of the temperature dependence of $K$ has been reported in Ref.~\onlinecite{Cui2015}, we are not going to discuss it in this paper. 
    In order to extract $K_{\text{spin}}(T)$, which is needed for the following Korringa ratio analysis, we plot $K(T)$ against the corresponding bulk static uniform magnetic susceptibility $\chi(T)$ with $T$ as an implicit parameter as shown in Fig. \ref{fig:T-K}(b).
    From the slope of the linear fit curve,  the hyperfine coupling constant can be estimated. 
   The $x$ dependence of the hyperfine coupling constant has been reported in Ref. \onlinecite{Cui2015}.  
    From the $y$-intercept of the linear fit curve, one can estimate the orbital shift $K_0$, and extract $K_{\text{spin}}(T)$  to discuss magnetic correlations.

      A  Korringa ratio analysis is applied to extract the character of spin fluctuations in Ca(Fe$_{1-x}$Co$_x$)$_2$As$_2$  from $^{75}$As NMR data as has been carried out for both the electron-doped Ba(Fe$_{1-x}$Co$_x$)$_2$As$_2$ and hole-doped Ba$_{1-x}$K$_x$Fe$_2$As$_2$ families of iron-pnictide SCs.\cite{Wiecki2015prl} 
      Within a Fermi liquid picture,  $1/T_1T$ is proportional to the square of the density of states ${\cal D}(E_{\rm F})$ at the Fermi energy and $K_{\text{spin}} (\propto \chi_{\text{spin}}$) is proportional to ${\cal D}(E_{\rm F})$. 
     In particular, $T_1TK_{\text{spin}}^2$  = $\frac{\hbar}{4\pi k_{\rm B}} \left(\frac{\gamma_{\rm e}}{\gamma_{\rm N}}\right)^2$ = ${\cal S}$, which is the Korringa relation.  
    For the $^{75}$As nucleus ($\gamma_{\rm N}/2\pi=7.2919$ MHz/T),  ${\cal S} =8.97\times 10^{-6}$ Ks. 
    Korringa ratio $\alpha\equiv$ ${\cal S}$$/(T_1TK_{\text{spin}}^2)$, which reflects the deviations from ${\cal S}$, can reveal information about how electrons correlate in the material.\cite{Moriya1963,Narath1968}
    $\alpha\sim1$ represents the situation of uncorrelated electrons. 
   On the other hand, $\alpha >1$ indicates AFM correlations while $\alpha <1$ for FM correlations.
    These come from the enhancement of $\chi(\mathbf{q}\neq 0)$, which increases $1/T_1T$ but has little or no effect on $K_{\text{spin}}$, since the latter probes only the uniform $\chi(\mathbf{q} = 0)$.  
   Therefore, the predominant feature of magnetic correlations, whether AFM or FM, can be determined by the Korringa ratio $\alpha$. 

    To proceed with the Korringa ratio analysis, one needs to take the anisotropy of $K_\text{spin}$ and $1/T_1T$ into consideration. 
    $1/T_1$ picks up the hyperfine field fluctuations at the NMR Larmor frequency, $\omega_{\rm 0}$, perpendicular to the applied field according to $(1/T_1)_{H||i}=\gamma_{\rm N}^2\left[|H^{\rm hf}_j(\omega_{\rm 0})|^2+|H^{\rm hf}_k(\omega_{\rm 0})|^2\right]$,
where $(i,j,k)$ are mutually orthogonal directions and $|H^{\rm hf}_j(\omega_{\rm 0})|^2$ represents the power spectral density of the $j$-th component of the hyperfine magnetic field at the nuclear site. 
    Thus, defining $H^{\rm hf}_{ab}\equiv H^{\rm hf}_{a}=H^{\rm hf}_{b}$, which is appropriate for the tetragonal PM state, we have $(1/T_1)_{H||c}=2\gamma_{\rm N}^2|H^{\rm hf}_{ab}(\omega_{\rm 0})|^2\equiv 1/T_{1,\perp}$. 
    The Korringa parameter $\alpha_{\bot}\equiv {\cal S}/T_{1,\bot}TK_{\text{spin},ab}^2$ will then characterize fluctuations in the $ab$-plane component of the hyperfine field. 
   Similarly, we consider the quantity $1/T_{1,\|}\equiv2(1/T_1)_{H||ab}-(1/T_1)_{H||c}= 2\gamma_N^2|H^{\rm hf}_{c}(\omega_{\rm N})|^2$, since $(1/T_1)_{H||ab}=\gamma_N^2\left[|H^{\rm hf}_{ab}(\omega_{\rm N})|^2+|H^{\rm hf}_c(\omega_{\rm N})|^2\right]$. We then pair $K_{\text{spin},c}$ with $1/T_{1,\|}$, so that the Korringa parameter $\alpha_{\|}={\cal S}/T_{1,\|}TK_{\text{spin},c}^2$ characterizes fluctuations in the $c$-axis component of the hyperfine field.

\begin{figure}[tb]
\includegraphics[width=8.0cm]{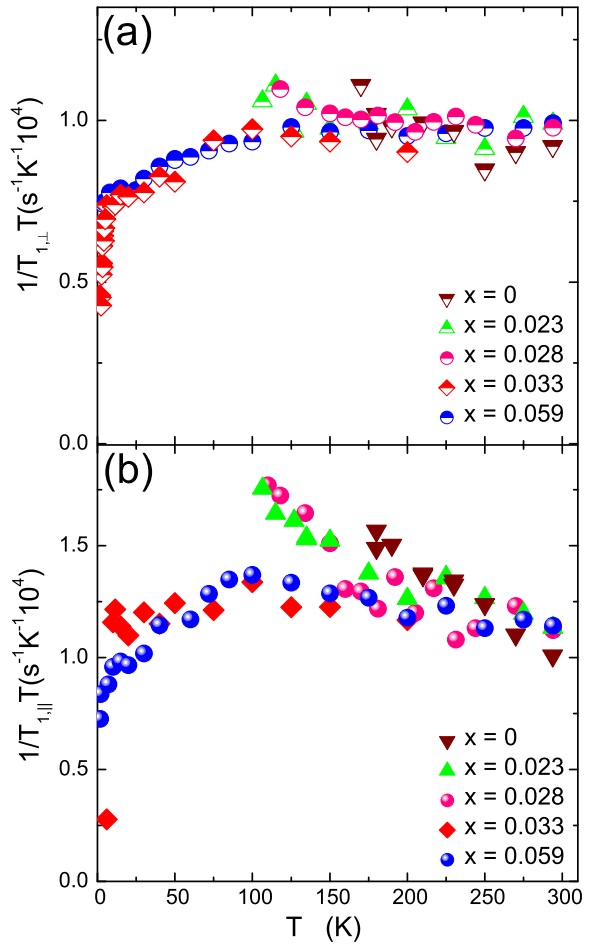} 
\caption{(Color online) Temperature dependence of 1/$T_1T$ with anisotropy in Ca(Fe$_{\rm 1-x}$Co$_{x}$)$_2$As$_2$. (a) $1/T_{1,\perp}$ = $(1/T_1T)_{H||c}$.   (b) $1/T_{1,\|}T=2(1/T_1T)_{H||ab}-(1/T_1T)_{H||c}$.}
 \label{fig:T1T}
\end{figure}

     Figure~\ref{fig:T1T} shows the temperature dependence of $1/T_{1,\perp}T$ and $1/T_{1,\|}T$  in Ca(Fe$_{1-x}$Co$_{x}$)$_2$As$_2$ at $H$ $\sim$ 7.5 T, obtained from  the $(1/T_1T)_{H||ab}$ and $ (1/T_1T)_{H||c}$ data reported previously.\cite{Cui2015}
  For $x$ = 0, 0.023, and 0.028, $1/T_{1,\|}T$s show a monotonic increase with decreasing $T$ down to $T_{\rm N}$ = 170, 106, and 53 K for $x$ = 0, 0.023, 0.028, respectively, while $1/T_{1,\perp}T$s are nearly independent of $T$ although the slight increase can be seen near $T_{\rm N}$ for each sample.
    Since the increase of $1/T_{1,\|}T$s originates from the growth of the stripe-type AFM spin fluctuations,\cite{Cui2015} the results indicate that the AFM spin fluctuations enhance the hyperfine fluctuations at the As sites along the $c$ axis.  
     In the case of superconducting samples with $x$ $\geq$ 0.033, $1/T_{1,\perp}T$ and $1/T_{1,\|}T$ show a slight increase or constant above $T^{*}$ $\sim$ 100 K on cooling and then start to decrease below $T^{*}$.
   These behaviors are  ascribed to pseudogap-like behavior in Ref. \onlinecite{Cui2015}.
    With a further decrease in $T$, both $1/T_{1,\|}T$ and $1/T_{1,\perp}T$  for $x$ = 0.033 and 0.059 show  sudden decreases below $T_{\rm c}$ [15 (10) K for $x$ =  0.033 (0.059)] due to superconducting transitions.

\begin{figure*}[tb]
\includegraphics[width=16.0cm]{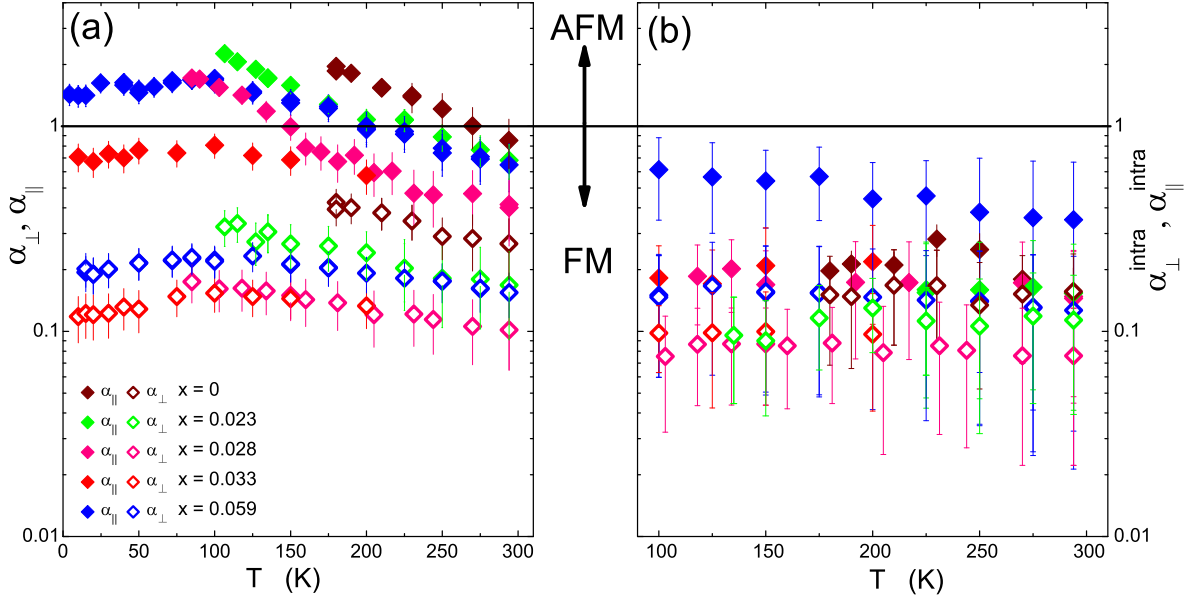} 
 \caption{(Color online) (a) $T$ dependence of Korringa ratios,  $\alpha_\perp$ and $\alpha_{\|}$. (b) $T$ dependence of intraband Korringa ratios, $\alpha_\perp^\text{intra}$and $\alpha_\|^\text{intra}$, above $T_{\rm N}$ or $T^*$, obtained by subtracting a CW term from the temperature dependence of $1/T_{1,\perp}T$ and $1/T_{1,\|}T$  as described in the text.
 }
 \label{fig:alpha}
 \end{figure*} 	

    Using the $1/T_{1,\perp}T$, $1/T_{1,\|}T$ data and Knight shift data, we discuss magnetic correlations in Ca(Fe$_{1-x}$Co$_{x}$)$_2$As$_2$ based on the Korringa ratios. 
   The $T$ dependences of the Korringa ratios $\alpha_{\bot}={\cal S}/T_{1,\bot}TK_{\text{spin},ab}^2$ and $\alpha_{\|}={\cal S}/T_{1,\|}TK_{\text{spin},c}^2$ are shown in Fig. \ref{fig:alpha}(a).
   All $\alpha_{\|}$ and $\alpha_\perp$ increase with decreasing $T$ down to $T_{\rm N}$ or $T^{*}$. 
   The increase in $\alpha$, which is the increase in $1/T_{1,}TK^2$, clearly indicates the growth of the stripe-type AFM spin correlations as have been pointed out previously.\cite{Cui2015} 
    It is noted that $\alpha_{\|}$ is always greater than  $\alpha_\perp$ for each sample, indicating that stronger hyperfine fluctuations at the As sites due to AFM correlations along the $c$ axis than in $ab$.
    On the other hand, $\alpha_{\|}$ values seem to be less than unity: the largest value of $\alpha_\perp$ can be found to be  $\sim$ 0.4 in $x$ = 0. 
    The even smaller values $\alpha_\perp$ of 0.1 -- 0.2 in $x$ = 0.023 and $x$ = 0.028 at high temperatures are observed, suggesting FM fluctuations in the normal state.

     In the application of the Korringa ratio to the iron pnictides, the question arises as to the role of the hyperfine form factor, which can, in principle, filter out the AFM fluctuations at the As site. 
    This filtering effect could affect the balance of FM vs.  AFM fluctuations as measured by the Korringa ratio. \cite{Jeglic2010}
   In order to discuss the filtering effects, it is convenient to express 1/$T_1$ in terms of wave-number ($\mathbf{q}$) dependent form factors and $\mathbf{q}$ dependent dynamical spin susceptibility $\chi(\mathbf{q}, \omega_0)$.  
    By an explicit calculation of the form factors (see Appendix A) using the methods of Ref. \onlinecite{Smerald2011},
we find that
\begin{equation}
\frac{1}{T_{1,\|}T}\sim
\left[\left(2.7\frac{{\rm T}^2}{\mu_{\rm B}^2}\right)\frac{\chi_{ab}''(\mathbf{Q},\omega_0)}{\hbar\omega_0}
+\left(1.5\frac{{\rm T}^2}{\mu_{\rm B}^2}\right)\frac{\chi_{c}''(\mathbf{0},\omega_0)}{\hbar\omega_0}\right],
\end{equation}
\begin{equation}
\frac{1}{T_{1,\perp}T}\sim
\left[\left(3.2\frac{{\rm T}^2}{\mu_{\rm B}^2}\right)\frac{\chi_{ab}''(\mathbf{0},\omega_0)}{\hbar\omega_0}
+\left(1.4\frac{{\rm T}^2}{\mu_{\rm B}^2}\right)\frac{\chi_{c}''(\mathbf{Q},\omega_0)}{\hbar\omega_0}\right]
\end{equation}
where $\chi''(\mathbf{0},\omega_0)$ and $\chi''(\mathbf{Q},\omega_0)$ represent the imaginary part of the dynamical susceptibility for $\mathbf{q}$ = 0 ferromagnetic and  $\mathbf{Q}$ = $(\pi,0)/(0,\pi)$  stripe-type AFM components, respectively. 
The numbers are calculated from the hyperfine coupling constants in units of T/$\mu_{\rm B}$ for CaFe$_2$As$_2$ given in Ref. \onlinecite{Cui2015}.
From these equations, it is clear that  the stripe-type AFM fluctuations are not filtered out for both directions in the iron pnictides.
    It is also seen that  for $1/T_{1,\|}T$, the form factor favors AFM fluctuations, which explains the larger (more AFM) values of $\alpha_\|$. 
   On the other hand, for $1/T_{1\perp}T$, the ferromagnetic fluctuations dominate more than the AFM fluctuations as actually seen in Fig. \ref{fig:alpha}(a) where $\alpha_\perp$ is less than $\alpha_\|$ for each sample. 

\begin{figure}[tb]
\includegraphics[width=8.0cm]{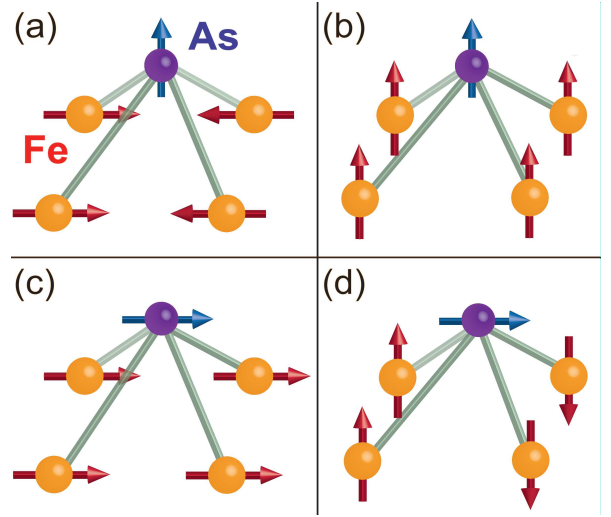} 
\caption{(Color online) 
(a),(b): Sources of hyperfine field along the $c$-axis. (c),(d): Sources of hyperfine field in the $ab$-plane.}
 \label{fig:hyperfine}
 \end{figure} 

\begin{figure*}[tb]
\includegraphics[width=17.0 cm]{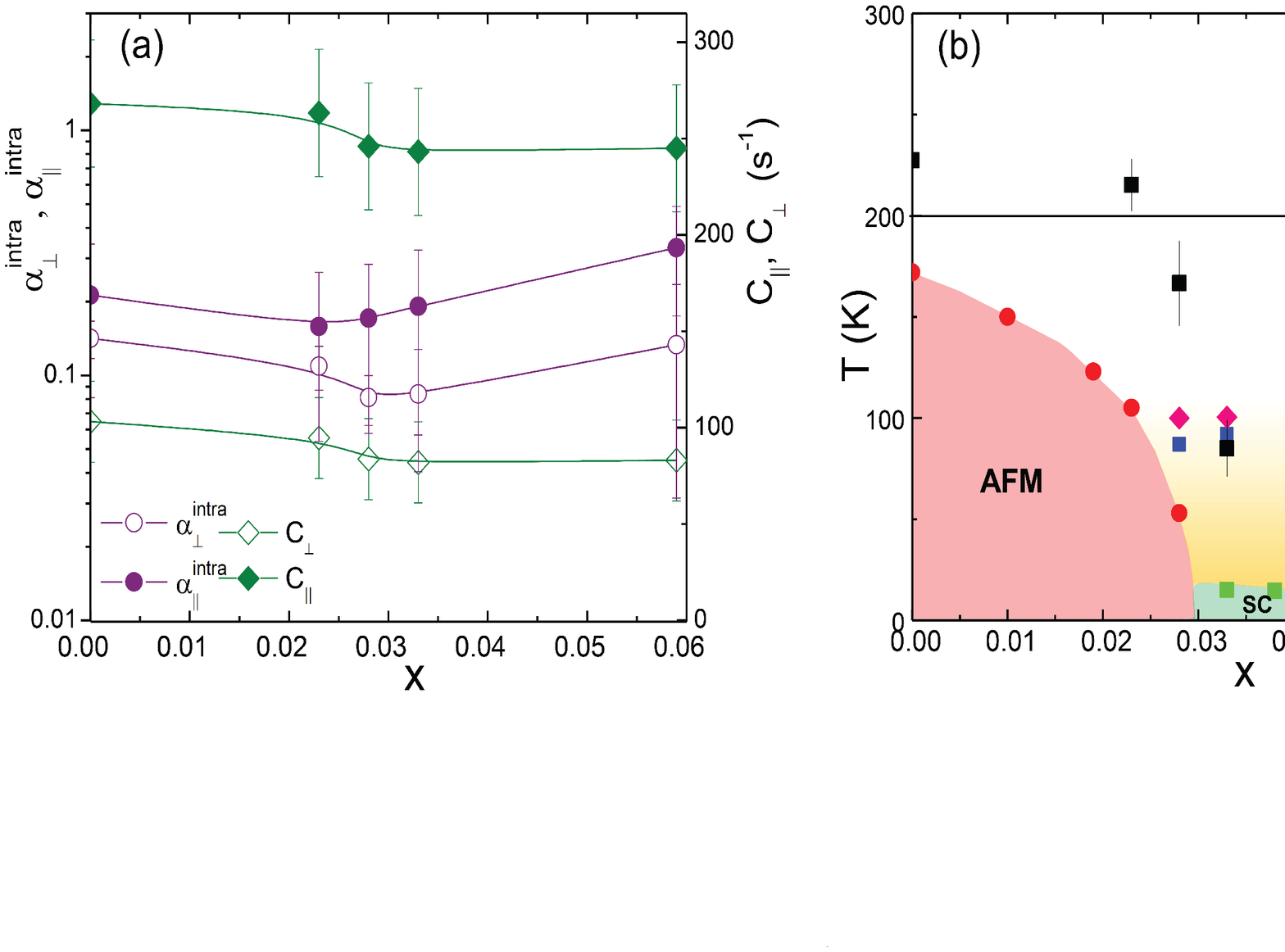} 
\caption{(Color online) (a) Doping dependence of the $T$-independent values of $\alpha_\perp^\text{intra}$, $\alpha_\|^\text{intra}$ and Curie-Weiss parameters  $C_\perp$, $C_\|$  
  The lines are guide for eyes. 
(b) Phase diagram of Ca(Fe$_{1-x}$Co$_x$)$_2$As$_2$. 
$T_{\rm N}$ and $T_{\rm c}$ are from Ref. \onlinecite{Ran2012}.
The pseudogap crossover temperature $T_{ab}^*$ and $T_{c}^*$ are determined by NMR measurements for $H$  $\parallel$ $ab$ plane and  $H$  $\parallel$ $c$ axis, respectively. 
AFM, SC and PG stand for the antiferromagnetic ordered state, superconducting, and pseudogap phases, respectively.  
}
\label{fig:phase}
\end{figure*}

      Now we consider the origin of the hyperfine field at the $^{75}$As site in order to further understand the physics associated with each term in Eqs. (1) and (2).  
    The hyperfine field at the $^{75}$As site is determined by the spin moments on the Fe sites through the hyperfine coupling tensor $\tilde{A}$, according to $\mathbf{H}^{\text{hf}}=\tilde{A}\cdot\mathbf{S}$.  In the tetragonal PM phase,  the most general form for $\tilde{A}$ is \cite{Kitagawa2008,Hirano2012}
\begin{equation}
\tilde{A}=
\begin{pmatrix}
  A_\perp & D & B \\
  D & A_\perp & B \\
  B & B & A_c
 \end{pmatrix},
\end{equation}
where $A_i$ is the coupling for FM correlation, $D$ is the coupling for in-plane Ne\'{e}l-type AFM correlation and $B$ is coupling for stripe-type AFM correlations. Since there is no theoretical or experimental reason to  expect Ne\'{e}l-type AFM correlation in the iron pnictides, below we simply set $D=0$.  
    We then obtain $H^\text{hf}_\perp=A_\perp S_\perp+BS_{\rm c}$ and $H^\text{hf}_{\rm c}=2BS_\perp+A_{\rm c}S_{\rm c}$. 
   There are therefore two sources of hyperfine field pointing along the $c$ axis\cite{Kitagawa2008}: fluctuations at $\mathbf{q}=\mathbf{Q}=(\pi,0)/(0,\pi)$ with the spins pointing in plane (as illustrated in Fig. \ref{fig:hyperfine}(a)) or fluctuations at $\mathbf{q}=0$ with the spins pointing along the $c$ axis (Fig. \ref{fig:hyperfine}(b)). 
   The first and second fluctuations correspond to the first and second terms, respectively, in $1/T_{1, \|}T$ [Eq. (1)]. 
   Similarly, hyperfine field fluctuations in the $ab$ plane can result from fluctuations at $\mathbf{q}=0$ with the spins pointing in plane (Fig. \ref{fig:hyperfine}(c)), or from fluctuations at $\mathbf{q}=\mathbf{Q}$ with the spins pointing along the $c$ axis (Fig. \ref{fig:hyperfine}(d)). 
    Again, the first and second fluctuations can be attributed to the first and second terms, respectively, in $1/T_{1, \perp}T$ [Eq. (2)]. 
   In what follows, we will refer to the correlations depicted in Fig. \ref{fig:hyperfine}(a) as ``(a)-type'' correlations (similarly for the others).
   To summarize, the value of $\alpha_\|$ reflects the competition between (a)- and (b)-type correlations, while 
$\alpha_\perp$ reflects the competition between (c)- and (d)-type correlations.

  Now, since $\alpha_\|$ reflects the character of hyperfine field fluctuations with a $c$-axis component, the strongly AFM $\alpha_\|$ in Fig. \ref{fig:alpha} can be attributed to  stripe-type AFM correlations with the Fe spins in plane (i.e. (a)-type). 
   These must dominate the (b)-type correlations in order to have an AFM value of $\alpha_\|$.
    Similarly, since $\alpha_\perp$ reflects the character of the $ab$- plane component of hyperfine field fluctuations, the strongly FM value of $\alpha_\perp$ in the high $T$ region may be attributed to in plane FM fluctuations (Fig. \ref{fig:hyperfine}(c)), while the increase of $\alpha_\perp$ as the temperature is lowered reflects the increasing dominance of stripe-type AFM correlations with a $c$-axis component to the spin (as in Fig. \ref{fig:hyperfine}(d)). 
   By examining the $c$-axis and $ab$-plane components of the hyperfine field fluctuations separately via $\alpha_\|$ and $\alpha_\perp$, we see the simultaneous coexistence of FM and AFM fluctuations in Ca(Fe$_{1-x}$Co$_x$)$_2$As$_2$.
    Furthermore, the dominance of  (a)- and (c)-type spin fluctuations in the high temperature region suggests that both the AFM and FM fluctuations are highly anisotropic, favoring the $ab$-plane.
   A similar feature of the coexistence of FM and AFM fluctuations\cite{Wiecki2015prl}  has been reported in Ba(Fe$_{1-x}$Co$_x$)$_2$As$_2$ and Ba$_{1-x}$K$_x$Fe$_2$As$_2$.

    It is interesting  to separate the FM and the stripe-type AFM fluctuations and extract their $T$ dependence, as has been performed in the hole- and electron-doped BaFe$_2$As$_2$.\cite{Wiecki2015prl}
     According to the previous paper,\cite{Wiecki2015prl}  $1/T_1T$ was decomposed into  inter- and intraband components according to $1/T_1T=(1/T_1T)_\text{inter}+(1/T_1T)_\text{intra}$, where the $T$ dependence of the interband term is assumed to follow the Curie-Weiss (CW) form appropriate for 2D AFM spin fluctuations: $(1/T_1T)_\text{inter}=C/(T-\Theta_{\rm CW})$.
  For $T$ dependence of the intraband component, $(1/T_1T)_\text{intra}$ was assumed to be $(1/T_1T)_\text{intra}$ = $\alpha+\beta\text{exp}(-\Delta/k_{\rm B}T)$.  
     Here we also tried to decompose the present $1/T_{1,\|}T$ and $1/T_{1,\perp}T$ data following the procedure.
We, however,  found large uncertainty in decomposing our data, especially for the $1/T_{1,\perp}T$ case, due to the weak temperature dependence of 1/$T_1T$.      
   Nevertheless, we proceeded with our analysis to qualitatively examine the $x$ dependence of Curie-Weiss parameter $C$, which measures the strength of AFM spin fluctuations, and  $\Theta_{\rm CW}$ corresponding to the distance in $T$ from the AFM instability point.
   Here we fit the data above $T_{\rm N}$ or $T^*$ for each sample.
    $\Theta_{\rm CW}$  decreases from  38 $\pm$ 17 K ($x$ = 0) to 15 $\pm$ 13 K ($x$ = 0.023), and to a negative values of --33 $\pm$ 21 K ($x$ = 0.028). 
 This suggests  that compounds with $x$ = 0.023 and 0.028 are close to the AFM instability point of $\Theta_{\rm CW}$  = 0 K.
  A similar behavior of $\Theta_{\rm CW}$  is reported in  Ba(Fe$_{1-x}$Co$_x$)$_2$As$_2$ (Refs. \onlinecite{Wiecki2015prl,Ning2010}) and Ba(Fe$_{1-x}$Ni$_x$)$_2$As$_2$ (Ref. \onlinecite{Guoqing}). 
     The $x$ dependences of CW parameters $C_\perp$, $C_\|$  and $\Theta_{\rm CW}$ are shown in Figs. \ref{fig:phase}(a) and (b)  together with the phase diagram reported in Ref. \onlinecite{Cui2015}.   
    Although these parameters have large uncertainty,   $C_\|$ seems to be greater than $C_\perp$, consistent with that the in-plane AFM fluctuations are stronger than the $c$-axis AFM fluctuations. 
   This result is same as in Ba(Fe$_{1-x}$Co$_{x}$)$_2$As$_2$ samples in Ref. \onlinecite{Wiecki2015prl}. 
  On the other hand, the $C_\perp$ and $C_\|$ parameters are almost independent of $x$ in Ca(Fe$_{1-x}$Co$_{x}$)$_2$As$_2$ in the substitution range of $x$ = 0--0.059,  while the $C_\perp$ and $C_\|$ parameters decrease with Co substitution in BaFe$_2$As$_2$ where the $c$-axis component AFM spin fluctuations decrease and die out with $x$ $\geq$ 0.15.\cite{Ning2010} 
   It is interesting to point out that a similar $x$-independent behavior is also observed in the crossover temperature $T^*$ attributed to the pseudogaplike behavior in the spin excitation spectra of Ca(Fe$_{1-x}$Co$_{x}$)$_2$As$_2$ system.\cite{Cui2015} 


    Finally we show, in Fig.~\ref{fig:alpha}(b), the intra band Korringa ratios $\alpha_\|^\text{intra}$ and $\alpha_\perp^\text{intra}$ by subtracting the  interband scattering term $C$/$(T-\Theta_{\rm CW})$.
   Both $\alpha_\|^\text{intra}$ and $\alpha_\perp^\text{intra}$ remain roughly constant above $T_{\rm N}$ or $T^*$. 
   We plotted the average value of  $\alpha_\|^\text{intra}$ and $\alpha_\perp^\text{intra}$ as a function of $x$ in Fig.~\ref{fig:phase}(b). 
      We find that $\alpha_\perp^\text{intra}$ is smaller than $\alpha_\|^\text{intra}$ for all the samples, confirming again the dominant in-plane FM spin fluctuations. 
     The calculated $\alpha_\perp^\text{intra}$and $\alpha_\|^\text{intra}$ in Ca(Fe$_{1-x}$Co$_{x}$)$_2$As$_2$ are almost same order with those in both the electron and hole doped BaFe$_2$As$_2$.
     These results indicate that the FM spin correlations exist in general  and may be a key ingredient to a theory of superconductivity in the iron pnictides.

 \section{Summary}  

    Motivated by the recent NMR measurements which revealed the coexistence of the stripe-type antiferromagnetic (AFM) and ferromagnetic (FM) spin correlations in both the hole- and electron-doped BaFe$_2$As$_2$ families of iron-pnictide superconductors\cite{Wiecki2015prl},  we have reanalyzed NMR data in  Ca(Fe$_{1-x}$Co$_x$)$_2$As$_2$ and found clear evidence for the coexistence of the stripe-type AFM  and FM spin correlations. 
       In contrast to the case of Ba(Fe$_{1-x}$Co$_x$)$_2$As$_2$ where the relative strength of FM correlations increases with Co substitution,  the FM correlations are almost independent of the Co substitution for our investigated range of $x$ = 0 -- 0.059 in Ca(Fe$_{1-x}$Co$_x$)$_2$As$_2$. 
        The Curie-Weiss parameters $C_{\perp,\|}$ representing the strength of the stripe-type AFM correlations are almost independent of the Co doping, close to a feature of $T^*$ representing a characteristic  temperature of the pseudogaplike behavior. 
    Our analysis of the NMR data indicates that FM fluctuations exist in general in iron-pnictide superconducting families.  
    Further systematic theoretical and experimental investigation on the role of the FM correlations in iron pnictide superconducting families are highly required.

 \section{Acknowledgments}
       We thank David C. Johnston for helpful discussions. 
       The research was supported by the U.S. Department of Energy, Office of Basic Energy Sciences, Division of Materials Sciences and Engineering. Ames Laboratory is operated for the U.S. Department of Energy by Iowa State University under Contract No.~DE-AC02-07CH11358.

\appendix
\section{A calculation of form factor}
Here, we directly calculate the appropriate form factors for the PM state of the iron pnictides according to the theory of Ref.~\onlinecite{Smerald2011}.
   We make the assumption that the external applied field is much larger than the hyperfine field, which is certainly true in the PM state. 
    We further assume that the wave-number $q$ dependent dynamic susceptibility tensor $\chi^{\alpha\beta}(\mathbf{q},\omega_0)$ is diagonal in the PM state. 
    Under these assumptions, the spin-lattice relaxation rate in an external field $\mathbf{h}_{\rm{ext}}$ is given by
\begin{equation}
\frac{1}{T_1(\mathbf{h}_{\rm{ext}})}=\lim_{\omega_0 \to 0}\frac{\gamma_N^2}{2N}k_{\rm B}T
\sum_{\alpha,\mathbf{q}}{\cal{F}}_\alpha^{\mathbf{h}_{\rm{ext}}}(\mathbf{q})
\frac{\rm{Im}[\chi^{\alpha\alpha}(\mathbf{q},\omega_0)]}{\hbar\omega_0},
\label{eq:T1}
\end{equation}
where $\alpha=(a,b,c)$ sums over the crystallographic axes. The general expression for the $q$ dependent 
form factor is 
\begin{equation}
{\cal{F}}_\alpha^{\mathbf{h}_{\rm{ext}}}(\mathbf{q})=\sum_{\gamma,\delta}
[R_{\mathbf{h}_{\rm{ext}}}^{x\gamma}R_{\mathbf{h}_{\rm{ext}}}^{x\delta}+(x\leftrightarrow y)]
{\cal A}_\mathbf{q}^{\gamma\alpha}{\cal A}_{-\mathbf{q}}^{\delta\alpha},
\label{eq:form}
\end{equation}
where $R_{\mathbf{h}_{\rm{ext}}}$ is a matrix which rotates a vector from the crystallographic $(a,b,c)$
coordinate system to a coordinate system $(x,y,z)$ whose $z$ axis is 
aligned with the total magnetic field at the nuclear site. For details we refer the reader to 
Ref. \onlinecite{Smerald2011}. When $\mathbf{h}_{\rm{ext}}\|c$, the two coordinate systems coincide so that
\begin{equation}
R_{\mathbf{h}_{\rm{ext}}\|c}=
\begin{pmatrix}
  1 & 0 & 0 \\
  0 & 1 & 0 \\
  0 & 0 & 1
\end{pmatrix}.
\label{eq:rc}
\end{equation}
For $\mathbf{h}_{\rm{ext}}\|a$, the appropriate matrix is
\begin{equation}
R_{\mathbf{h}_{\rm{ext}}\|a}=
\begin{pmatrix}
  0 & 0 & 1 \\
  0 & 1 & 0 \\
  -1 & 0 & 0
\end{pmatrix}.
\label{eq:ra}
\end{equation}
For the case of the As site in the iron pnictides, the matrix ${\cal A}_\mathbf{q}$ in Eq. \ref{eq:form}
is given by \cite{Smerald2011}
\begin{equation}
{\cal A}_\mathbf{q}=4
\begin{pmatrix}
  {\cal A}^{aa}c_ac_b & -{\cal A}^{ab}s_as_b & i{\cal A}^{ac}s_ac_b \\
  -{\cal A}^{ba}s_as_b & {\cal A}^{bb}c_ac_b & i{\cal A}^{bc}c_as_b \\
  i{\cal A}^{ca}s_ac_b & i{\cal A}^{cb}c_as_b & {\cal A}^{cc}c_ac_b
\end{pmatrix},
\label{eq:aq}
\end{equation}
where ${\cal A}^{\alpha\beta}$ are the components of the hyperfine coupling tensor and 
\begin{align*}
c_a&=\cos\frac{q_aa_0}{2}&c_b&=\cos\frac{q_bb_0}{2}\\
s_a&=\sin\frac{q_aa_0}{2}&s_b&=\sin\frac{q_bb_0}{2}.
\end{align*}
Here $a_0$ and $b_0$ are lattice constants. Of course, $a_0=b_0$ in the PM state.
Combining Eqs. \ref{eq:form}-\ref{eq:aq}, we obtain
\begin{align}
{\cal{F}}_a^{\mathbf{h}_{\rm{ext}}\|a}(\mathbf{q})&=16({\cal A}^{ca}s_ac_b)^2+16({\cal A}^{ba}s_as_b)^2\\
{\cal{F}}_b^{\mathbf{h}_{\rm{ext}}\|a}(\mathbf{q})&=16({\cal A}^{cb}c_as_b)^2+16({\cal A}^{bb}c_ac_b)^2\\
{\cal{F}}_c^{\mathbf{h}_{\rm{ext}}\|a}(\mathbf{q})&=16({\cal A}^{cc}c_ac_b)^2+16({\cal A}^{bc}c_as_b)^2
\end{align}
and
\begin{align}
{\cal{F}}_a^{\mathbf{h}_{\rm{ext}}\|c}(\mathbf{q})&=16({\cal A}^{aa}c_ac_b)^2+16({\cal A}^{ba}s_as_b)^2\\
{\cal{F}}_b^{\mathbf{h}_{\rm{ext}}\|c}(\mathbf{q})&=16({\cal A}^{bb}c_ac_b)^2+16({\cal A}^{ab}s_as_b)^2\\
{\cal{F}}_c^{\mathbf{h}_{\rm{ext}}\|c}(\mathbf{q})&=16({\cal A}^{ac}s_ac_b)^2+16({\cal A}^{bc}c_as_b)^2.
\end{align}
To calculate $1/T_1$ from Eq. \ref{eq:T1}, we assume for simplicity that 
$\chi^{\alpha\beta}(\mathbf{q},\omega_0)$ is non-zero only near the wavevectors $\mathbf{q}=0$, 
$\mathbf{q}=\mathbf{Q}_a\equiv(\pm\pi/a_0,0)$ and $\mathbf{q}=\mathbf{Q}_b\equiv(0,\pm\pi/b_0)$. 
By tetragonal symmetry we have $a\leftrightarrow b$. In particular,  $\mathbf{Q}_a=\mathbf{Q}_b\equiv\mathbf{Q}$ and $\rm{Im}[\chi^{aa}(\mathbf{q},\omega_0)]=\rm{Im}[\chi^{bb}(\mathbf{q},\omega_0)] \equiv\chi_{ab}''(\mathbf{q},\omega_0)$.
We also now write 
$\rm{Im}[\chi^{cc}(\mathbf{q},\omega_0)] \equiv\chi_{c}''(\mathbf{q},\omega_0)$.
We thus obtain


\begin{align}
\frac{1}{T_1(\mathbf{h}_{\rm{ext}}\|c)} = \lim_{\omega_0 \to 0}&\frac{8\gamma_N^2}{N}k_{\rm B}T
\left[2({\cal A}^{aa})^2\frac{\chi_{ab}''(\mathbf{0},\omega_0)}{\hbar\omega_0}\right. \nonumber\\
&\qquad \left.+ 4({\cal A}^{ac})^2\frac{\chi_{c}''(\mathbf{Q},\omega_0)}{\hbar\omega_0}\right]
\label{eq:T1c}
\end{align}

and
 \begin{align}
\frac{1}{T_1(\mathbf{h}_{\rm{ext}}\|a)}&=\lim_{\omega_0 \to 0}\frac{8\gamma_N^2}{N}k_{\rm B}T  
\left[
4({\cal A}^{ca})^2\frac{\chi_{ab}''(\mathbf{Q},\omega_0)}{\hbar\omega_0}\right.  \nonumber\\
&\qquad \left.+({\cal A}^{aa})^2\frac{\chi_{ab}''(\mathbf{0},\omega_0)}{\hbar\omega_0}\right. \nonumber\\
& \qquad \left.+({\cal A}^{cc})^2\frac{\chi_{c}''(\mathbf{0},\omega_0)}{\hbar\omega_0}\right.  \nonumber\\
&\qquad \left.+2({\cal A}^{ac})^2\frac{\chi_{c}''(\mathbf{Q},\omega_0)}{\hbar\omega_0}
\right].
\label{eq:T1a}
\end{align}
We have summed over four AFM wavevectors $\mathbf{Q}=(\pm\pi/a_0,0)$ and $\mathbf{Q}=(0,\pm\pi/a_0)$,
which have the same value of $\chi''(\mathbf{Q},\omega_0)$ in the PM state.
Notice that, for both field directions, AFM flucutations at $\mathbf{q}=\mathbf{Q}$ are completely filtered
out if ${\cal A}^{ac}=0$, as pointed out in Ref. \onlinecite{Johnston2010}.
However, in the iron pnictides ${\cal A}^{ac}\neq0$, \cite{Kitagawa2008}
and therefore AFM fluctuations are not filtered out.
From Eqs. \ref{eq:T1c} and \ref{eq:T1a}
we can easily calculate $1/T_{1,\|}\equiv2/T_1(\mathbf{h}_{\rm{ext}}\|a)-1/T_1(\mathbf{h}_{\rm{ext}}\|c)$ 
and $1/T_{1,\perp}\equiv1/T_1(\mathbf{h}_{\rm{ext}}\|c)$
\begin{align}
\frac{1}{T_{1,\perp}}=\lim_{\omega_0 \to 0}&\frac{16\gamma_N^2}{N}k_BT
\left[({\cal A}^{aa})^2\frac{\chi_{ab}''(\mathbf{0},\omega_0)}{\hbar\omega_0}\right. \nonumber\\
& \qquad \left.+2({\cal A}^{ac})^2\frac{\chi_{c}''(\mathbf{Q},\omega_0)}{\hbar\omega_0}\right]
\end{align}
\begin{align}
\frac{1}{T_{1,\|}}=\lim_{\omega_0 \to 0}&\frac{16\gamma_N^2}{N}k_BT \left[4({\cal A}^{ca})^2\frac{\chi_{ab}''(\mathbf{Q},\omega_0)}{\hbar\omega_0}\right. \nonumber\\
& \qquad \left.+ ({\cal A}^{cc})^2\frac{\chi_{c}''(\mathbf{0},\omega_0)}{\hbar\omega_0}\right]
\end{align}
Notice that the fluctuations probed by $1/T_{1,\|}$ and $1/T_{1,\perp}$ are consistent
with the qualitative arguments used in the main text. 
For the case of CaFe$_2$As$_2$, Ref. \onlinecite{Cui2015} gives ${\cal A}^{aa}=1.8$ T$/\mu_{\rm B}$,
${\cal A}^{cc}=1.2$ T$/\mu_{\rm B}$ and 
${\cal A}^{ca}={\cal A}^{ac}=0.82$ T$/\mu_{\rm B}$.
${\cal A}^{aa}$ and ${\cal A}^{cc}$ are determined by Knight shift measurements and 
${\cal A}^{ac}$ is found by comparing the measured internal field in the AFM state to the value of the 
ordered moment obtained by neutron scattering.

\bigskip*\ present address: Department of Physics, University of California, San Diego. California  92093, USA

\end{document}